# Coupling between waveguides and microresonators: the local approach

**DASHIELL L. P. VITULLO,**[1,*] **SAJID ZAKI,**[2] **D. E. JONES,**[1] **M. SUMETSKY,**[2] **AND MICHAEL BRODSKY**[1,3]

[1]*U.S. Army Research Laboratory, Adelphi, MD 20783-1193, USA*
[2]*Aston Institute of Photonic Technologies, Aston University, Birmingham B4 7ET, UK*
[3]*U.S. Military Academy, West Point, NY 10996, USA*
*\*Dashiell.L.Vitullo.civ@mail.mil*

**Abstract:** Coupling between optical microresonators and waveguides is a critical characteristic of resonant photonic devices with complex behavior that is not well understood. When the characteristic variation length of the microresonator modes is much larger than the waveguide width, local coupling parameters emerge that are independent of the resonator mode distributions and offer a simplified description of coupling behavior. We develop a robust numerical-fitting-based methodology for experimental determination of the local coupling parameters in all coupling regimes and demonstrate their characterization along a microfiber waveguide coupled to an elongated bottle microresonator.

## 1. Introduction

Photonic devices based on optical microresonators typically include waveguides, which are used to couple light in and out of microresonators. The performance of these devices is determined by the intrinsic optical characteristics of the microresonators and waveguides as well as by the coupling between them. The theoretical and experimental investigation of microresonators with different shapes (rings, spheres, toroids, bottles, etc.) is of great current interest and has been intensively developed for different applications [1-3]. While recent studies have identified promising novel coupling designs [4-6], less attention has been given to investigating exactly how coupling performance depends upon the optical and geometric characteristics of waveguides and microresonators [7-13]. These dependencies are quite complex [7,8] and in many cases it is easier to determine them experimentally [9-14]. However, understanding the fundamental features of coupling between waveguides and microresonators, especially of those with three-dimensional geometry (e.g., microspheres and microbottles), is important for the future development of resonant microdevices for classical [7,8,11-14] and quantum [9,10,15-17] applications.

Evanescent coupling between tapered fibers and whispering gallery modes (WGMs) is ultimately concerned with overlap integrals of the taper and resonator fields [7]. Typical coupling characterization focuses on quantities such as the transmission, roundtrip loss, and coupling strength [18], or ideality [10]. Determination of these parameters can indicate when parasitic losses are minimized, but does not provide details about the underlying loss processes. The local coupling approach, proposed in [19], applies in the regime where the characteristic length scale of the waveguide field $w$ is small compared to the transverse extent of the resonator fields with characteristic length $x_c$. When $w \ll x_c$, the waveguide-microresonator coupling is determined by the local value of the WGM microresonator field at the waveguide position, hence the name of the approach. This approximation simplifies the overlap integral, enabling separation of coupling parameters from the resonator field modes. Characterization of these parameters as the coupling configuration is varied enables insight into the underlying coupling and scattering processes. For example, resonant and non-resonant loss are described by separate parameters, thus yielding more insight into how to ameliorate loss than is afforded with a single loss parameter. The local coupling approach has been previously applied to the design of

Surface Nanoscale Axial Photonics (SNAP) devices, e.g. [20-22], but potential applications of this technology in the single photon regime require optimization of loss and coupling, which makes it critical to determine the dependencies of the coupling parameters upon transverse positioning in taper-microresonator systems.

In this article, we extend the local coupling approach with novel fitting capabilities that robustly determine the bare resonator modes and coupling parameters with quantified residual error and coupling parameter uncertainty estimates. We report the first characterization of the profile of these coupling parameters along the longitudinal axis of a tapered optical fiber. The procedure demonstrated herein maps the entire menu of coupling configurations available via transverse positioning of the taper along the longitudinal axis of the resonator with the two devices in contact, enabling subsequent selection of the desired coupling. Lastly, we report a novel quantification of the "criticality bound" that indicates how to determine the coupling regime (under-, critically, or over-coupled) from the coupling parameters.

## 2. Theoretical background

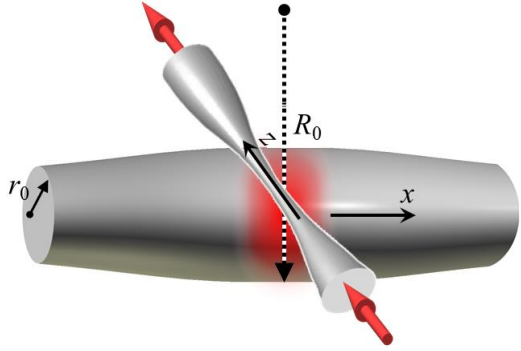

**Fig. 1.** WGM in a bottle microresonator. $R_0$ and $r_0$ are the axial and radial radii of curvature, respectively.

We investigate an elongated bottle microresonator coupled to a tapered optical fiber with micron-scale diameter (microfiber), as illustrated in Fig. 1. The fundamental WGM in this resonator behaves as $\exp[im\phi]\exp(-x^2/x_c^2)$ where $m$ is the azimuthal quantum number and $x_c = (2R_0 r_0)^{1/4} \lambda_{\text{res}}^{1/2} (2\pi n_e)^{-1/2}$ with resonance wavelength $\lambda_{\text{res}}$, effective refractive index $n_e$, and axial and radial radii of curvature $R_0$ and $r_0$, respectively [23]. Using $R_0 = 30$ m and $r_0 = 19$ μm, we have $x_c = 75$ μm which is significantly greater than the diameter of microfiber $w \sim 1$ μm used in our experiment, satisfying the local coupling condition.

We change the coupling by moving one of the devices relative to the other to vary the position where the microfiber and bottle microresonator make contact. The coupling depends on the local diameter of the taper at the contact point along its longitudinal axis z, as well as the position of the contact point along the resonator's longitudinal axis x. We set $z = 0$ when the resonator is aligned for contact in the center of the taper waist region, and $x = 0$ when the taper is aligned for contact with the center of the resonator. The range of the effective radius variation $\Delta r_{\text{eff}}(x)$ describing the bottle microresonator used in our experiment is very small (nanoscale); therefore, the resonant transmission power through the microfiber is described by [19]

$$P(\lambda, x, z) = \left| S_0(z) - \frac{i|C(z)|^2 G(\lambda,x,x)}{1+D(z)G(\lambda,x,x)} \right|^2, \quad (1)$$

where $\lambda$ is the vacuum wavelength. Here $S_0(z)$, $|C(z)|^2$ and $D(z)$ are the *local coupling parameters*, which depend on neither $x$ nor the cavity mode (interpretations detailed below). $G(\lambda, x, x)$ is the Green's function of the one-dimensional wave equation describing the propagation of WGMs along the bottle axis $x$:

$$\frac{\partial^2 \Psi}{\partial x^2} + \beta^2(\lambda, x)\Psi = 0. \tag{2}$$

Here $\beta(\lambda, x) = 2^{1/2} \beta_0 \left[\left(\frac{\Delta r_{\text{eff}}(x)}{r_0}\right) - \left(\frac{\Delta \lambda}{\lambda_{\text{res}}}\right)\right]^{1/2}$ is the WGM propagation constant, $\beta_0 = \frac{2\pi n}{\lambda_{\text{res}}}$ is the propagation constant in the bulk resonator material with refractive index $n$, and $\Delta \lambda$ is the wavelength variation [23].

The interpretations of the local coupling parameters in Eq. (1) are as follows: $|C|^2$ is the coupling strength between resonator and taper modes. $|S_0|$ describes the field transmission through the taper in the absence of coupling to resonator modes ($|C|^2 \to 0$). $|S_0|^2$ is the power transmission for light with nonresonant wavelength [where $G(\lambda, x, x) \approx 0$]. Transmission of resonant light depends on a coherent combination of the terms and exhibits Fano line-shapes, and the phase $\arg(S_0)$ controls the spectral shape of the resonances. The presence of the dielectric tapered fiber in the evanescent resonator field changes the field distribution relative to the condition where it is absent. $D$ describes these effects and relates the bare Green's function describing the resonator mode field in the absence of the taper $G(\lambda, x, x)$ to the renormalized (dressed) Green's function with the taper present $\bar{G}(\lambda, x, x) = \frac{G(\lambda, x, x)}{1 + D(z)G(\lambda, x, x)}$.

Re($D$) describes the shift of the resonance wavelength induced by the tapers presence (coupling to the taper changes the optical path length). Finally, Im(D) describes broadening of the resonances due to additional loss induced by the presence of the microfiber (e.g. via coupling to radiation modes). See [19] for additional background details.

## 3. Experimental characterization

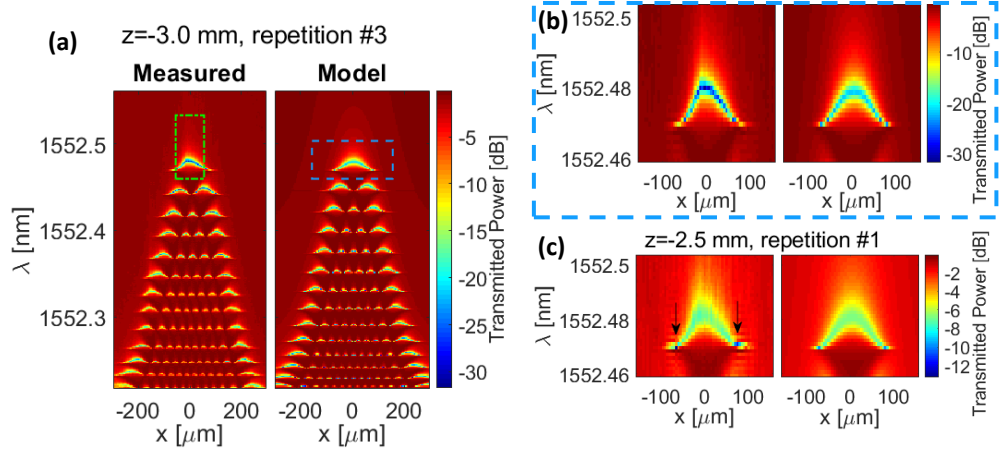

**Fig. 2.** (a) Comparison of measured and best-fit model spectrograms, near critical coupling, showing multiple axial modes. The green dot-dashed box over the measured data indicates the region used in coupling parameter fitting (see text). The blue dashed box over the model indicates the magnified region shown in (b) which compares the measured and best-fit model fundamental axial mode. (c) Comparison of measured and best-fit model fundamental resonances at $z = -2.5$ mm in the over-coupled regime. The characteristic edge dips seen in this regime are indicated with arrows on the measured data.

Our experimental system consists of an elongated SNAP bottle microresonator with ~400 $\mu$m extent along $x$, created on 38 $\mu$m diameter fiber using a $CO_2$ laser [24], coupled to a microfiber pulled using a ceramic microheater [25]. Coupling parameters are estimated through the measurement and analysis of 2D spectrograms, e.g. Fig. 2(a) [24]. Spectrograms are made by combining the transmission spectrum through the microfiber at multiple contact positions $x$ along the resonator with fixed $z$. The transmission spectrum is calculated from the Jones matrix spectrum of the system, measured with a Luna Technologies Optical Vector Analyzer (OVA). We isolate the Jones matrix describing transmission past the microresonator from those describing the taper segments and connecting fibers using the procedure described in [26]. From this, we calculate our reported transmission values, which are for light with polarization matched to the resonator modes. The baseline taper loss (spectral average of 4.6 dB) is removed such that transmitted power fraction is 0 dB (no loss) in the absence of coupling.

We then fit the measured spectrogram data to extract the best-fit coupling parameters. To accomplish this, we first find the Green's function solution [19] to the 1D wave equation of Eq. (2). The effective radius variation serves as a potential of the assumed form

$$\Delta r_{\text{eff}}(x) = A \exp\left[-\left(\frac{(x-x_0)^2}{2\,\sigma^2}\right)^p\right] + K. \tag{3}$$

We use a fitting procedure to find the values of $A, \sigma, p$ and $K$ that produce a modal eigenwavelength spectrum that best matches the observed spectrum. The best-fit values of $A = 3.2744$ nm, $\sigma = 123.5934$ $\mu$m, and $p = 1.1406$ are used for all resonators, while $x_0$ and $K$ are set for each spectrogram to account for the angle between the $x$ and $z$ axes being slightly different from 90°, and for random spectral shifts arising from thermal drift, respectively. Once the bare Green's functions $G(\lambda, x, x)$, for each spectrogram are found, the measured spectrograms are fit to Eq. (1) in the region indicated in the green box in Fig. 2(a) [see discussion below Eq. (5)] with fixed $G(\lambda, x, x)$ to find the 5 best-fit real-valued local coupling parameters: $|S_0(z)|^2$, $\arg[S_0(z)]$, $|C(z)|^2$, $\text{Re}[D(z)]$, and $\text{Im}[D(z)]$, in addition to the final minimized "cost" value (described below). Each spectrogram measurement is repeated 4 times to assess repeatability, and the profile of the mean average values for these parameters and the associated cost values (detailed below) are plotted in Fig. 3 with the error bars showing the standard deviation of each quantity.

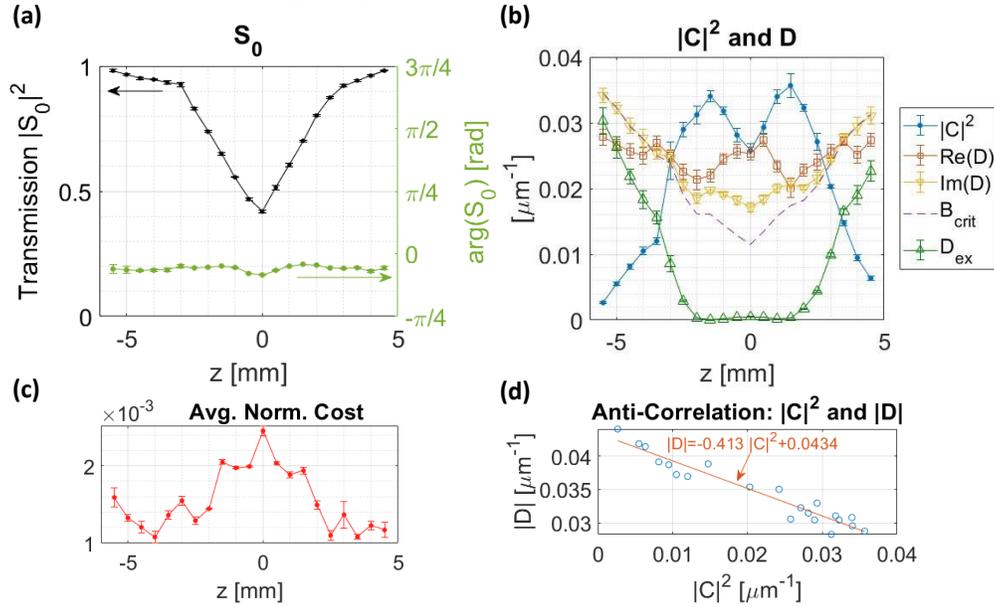

**Fig. 3**. Average coupling parameters with error bars showing standard deviation. $z$ indicates the position along the taper axis of resonator-taper contact, with $z = 0$ corresponding to resonator contact at the center of the taper waist region [25]. (a) Nonresonant transmission power amplitude $|S_0|^2$ with phase profile $\arg(S_0)$. (b) $|C|^2$ and $D$ parameters, with the critical coupling bound $B_{\text{crit}}$ [Eq. (5)] and the excess resonant loss $D_{\text{ex}}$ [Eq. (6)]. (c) The average cost value normalized as described near Eq. (4), indicates excellent agreement between model and theory. (d) The average values of $|C|^2$ and $|D|$ display anti-correlation. The best-fit line approximately describes the interesting relationship where stronger coupling is associated with smaller effect on the cavity $|D|$.

Characterization of the local coupling parameters yields rich information about coupling variation as the resonator is moved to vary the contact point along the taper axis $z$. The nonresonant power transmission $|S_0|^2$ [Fig. 3(a)] has slope transitions at $z = \pm 3$ mm and a mimimum value near the center of the taper waist region at $z = 0$ mm. The phase $\arg(S_0)$ is nearly flat across the entire measured $z$ range. Coupling strength $|C|^2$ peaks at $z = \pm 1.5$ mm [Fig. 3(b)]. $\text{Re}(D)$ has a roughly flat profile with random variation, indicating that the phase shift experienced by WGMs passing the microfiber is roughly independent of the microfiber diameter. The resonant loss $\text{Im}(D)$ is smallest in the central taper waist region, but increases with the local taper radius away from this region (as we discuss further with $D_{\text{ex}}$ below).

In some cases, the coupling parameter fits can converge to local minima that don't represent the actual coupling parameters. We determine when this occurs by comparing the best-fit model and experimental transmission amplitudes $S_{11}(\lambda, x, z) = S_0(z) - \frac{i\,|C(z)|^2\,G(\lambda,x,x)}{1+D(z)G(\lambda,x,x)}$. These two quantities are substantially different for local minima, and such a difference indicates that the fit must be run again with the local minima excluded, or with starting values closer to the true values.

The excellent agreement of our best-fit model and measured spectrograms is apparent from the low normalized cost [Fig. 3(c)]

$$\overline{\Delta P}(z) = \frac{\sqrt{\sum_{i,j}[P_{\text{meas}}(\lambda_i,x_j,z)-P_{\text{model}}(\lambda_i,x_j,z)]^2}}{H\,N}, \qquad (4)$$

where $P_{\text{meas}}$ and $P_{\text{model}}$ are the measured and best-fit model transmission [Eq. (1)], $i$ and $j$ index grid positions, the numerator is the cost value, and the denominator normalizes the cost by $N$, the number of transmission values in the fit region [green box in Fig. 2(a); the model has the same number of transmission values as the measured spectrograms], and $H$, the depth of the measured fundamental axial resonance along its central position (x=0). This quantifies the fractional variation per measured transmission value. The effectiveness of the local approach is validated by the small value of $\overline{\Delta P}(z)$ across the entire profile.

Microresonator-taper coupling can be sorted into three coupling regimes, set by the ratio of the light loss rate from the microresonator and the coupling rate between the microresonator and taper. Starting from the Fano formulation of transmission {Eq. (13) of [19]} and neglecting intrinsic material loss, it can be shown that the local coupling parameters determine the coupling regime. Critical coupling occurs when $|C(z)|^2$ equals the criticality bound

$$B_{\text{crit}}(z) = \frac{|S_0(z)|^2 \text{Im}[D(z)]}{\text{Re}[S_0(z)]}. \qquad (5)$$

The resonator-taper system is in the over-coupled regime when $|C(z)|^2 > B_{\text{crit}}(z)$, and in the undercoupled regime when $|C(z)|^2 < B_{\text{crit}}(z)$. This relationship also implies that the local group delay of individual axial modes is positive (negative) in the over-coupled (undercoupled) regime, c.f. Eq. (8) of [27].

Where $|z| > 3.0$ mm in Fig. 3(b), the system is undercoupled and $B_{\text{crit}}(z) \approx \text{Im}[D(z)]$. At $z = \pm 3.0$ mm, the power transmission for resonant light is very small (<2% for z=+3.0 mm; even smaller for z=-3.0 mm [Fig. 2]). This and the nearby crossings of $B_{\text{crit}}(z)$ and $|C(z)|^2$ both indicate that coupling is close to critical at these positions. Between these critical coupling positions, the system is over-coupled and it's important to perform the check described above against local minima. The transmission is very sensitive to small changes in over-coupled and critically-coupled configurations, and since our system uses no feedback stabilization, we note a concomitant increase in the standard deviation of $|C|^2$ in those regimes. We observe that the dips near the edges of the resonances in spectrograms [Fig. 2(c)] are indicative of over-coupling. The increased variation in these dips can confound the fit, which is why we select the fit-region indicated in Fig. 2(a).

The coupling parameters indicate device loss performance. Energy conservation sets two constraints on the coupling parameters [19]:

$$|S_0(z)| < 1 \quad \text{AND} \quad \text{Im}[D(z)] > |C(z)|^2 \frac{1-\text{Re}[S_0(z)]}{1-|S_0(z)|^2}, \tag{6}$$

which set bounds on the nonresonant and resonant loss, respectively. Minimum loss occurs when each of these conditions approaches equality. We quantify how much $\text{Im}[D(z)]$ exceeds this minimum with the excess resonant loss

$$D_{\text{ex}}(z) = \text{Im}[D(z)] - |C(z)|^2 \frac{1-\text{Re}[S_0(z)]}{1-|S_0(z)|^2}. \tag{7}$$

Investigation of the suggested proportionality relationship between the excess loss and the local radius of the microfiber at the point of contact is an interesting avenue for future research that could potentially be used to determine the microfiber radius variation (see e.g. [28]). We find a strong anti-correlation relationship between $|C|^2$ and $|D|$ (correlation coefficient = -0.96) [Fig. 3(d)], which indicates that the taper's effect on the cavity field (through resonant frequency shifts and induced loss) is smallest where the coupling is largest.

## 4. Conclusion

We report experimental characterization of the local coupling parameters, which describe the interaction between an elongated bottle microresonator and an input-output microfiber. In contrast to parameters commonly used for the description of the microresonator-waveguide coupling, these parameters are independent of the mode distribution. Our fitting approach demonstrates excellent agreement between measured and best-fit theoretical models, in addition to good coupling parameter repeatability between consecutive spectrogram measurements, in all coupling regimes (undercoupled through over-coupled). This method of characterizing coupling and loss paves the way for design optimization towards classical and quantum resonant optical devices. The elongated shape of the modes is of special importance since it allows us to simplify positioning of quantum emitters [29]. We suggest that, for this purpose, the microresonator profile can be optimized to arrive at enhanced regions with uniform WGM magnitude. Finally, we note that this approach can be generalized to find local coupling parameters with any microresonator system where $w \ll x_c$, through substitution of mode-solving methods appropriate to the resonator in use. Such generalization would enable investigation across multiple WGM resonator platforms to generate insight into commonalities and differences in their coupling behavior.

## Funding

## Disclosures

The authors declare no conflicts of interest.

## References

1. K. J. Vahala, "Optical Microcavities," Nature **424**, 839-846 (2003).
2. A. B. Matsko, *Practical Applications of Microresonators in Optics and Photonics* (CRC, 2009).
3. G. C. Righini, Y. Dumeige, P. Féron, M. Ferrari, G. Nunzi Conti, D. Ristic, and S. Soria, "Whispering gallery mode microresonators: Fundamentals and applications," Riv Nuovo Cimento **34**(7), 435-488 (2011).
4. X. Jiang, L. Shao, S.-X. Zhang, X. Yi, J. Wiersig, L. Wang, Q. Gong, M. Lončar, L. Yang, and Y.-F. Xiao, "Chaos-assisted broadband momentum transformation in optical microresonators," Science **358**(6361), 344-347 (2017).
5. F. Lei, J. M. Ward, P. Romagnoli, and S. Nic Chormaic, "Polarization-controlled cavity input-output relations," Phys. Rev. Lett. **124**(10) 103902 (2020).
6. N. Acharyya and G. Kozyreff, "Multiple Critical Couplings and Sensing in a Microresonator-Waveguide System," Phys. Rev. Appl. **8**(3) 034029 (2017).
7. B. E. Little, J.-P. Laine, and H. A. Haus, "Analytic Theory of Coupling from Tapered Fibers and Half-Blocks into Microsphere Resonators," J. Lightwave Technol. **17**(4), 704-715 (1999).
8. M. L. Gorodetsky and V. S. Ilchenko, "Optical microsphere resonators: optimal coupling to high-Q whispering-gallery modes," J. Opt. Soc. Am. B **16**(1), 147-154 (1999).

9. M. Cai, O. Painter, and K. J. Vahala, "Observation of Critical Coupling in a Fiber Taper to a Silica-Microsphere Whispering-Gallery Mode System," Phys. Rev. Lett. **85**(1), 74-77 (2000).
10. S. M. Spillane, T. J. Kippenberg, O. J. Painter, and K. J. Vahala, "Ideality in a Fiber-Taper-Coupled Microresonator System for Application to Cavity Quantum Electrodynamics," Phys. Rev. Lett. **91**, 043902 (2003).
11. M. J. Humphrey, E. Dale, A. T. Rosenberger, and D. K. Bandy, "Calculation of optimal fiber radius and whispering-gallery mode spectra for a fiber-coupled microsphere," Optics Commun. **271**(1), 124-131 (2007).
12. C.-L. Zou, Y. Yang, C.-H. Dong, Y.-F. Xiao, X.-W. Wu, Z.-F. Han, and G.-C. Guo, "Taper-microsphere coupling with numerical calculation of coupled-mode theory," J. Opt. Soc. Am. B **25**(11), 1895-1898 (2008).
13. A. Chiasera, Y. Dumeige, P. Féron, M. Ferrari, Y. Jestin, G. Nunzi Conti, S. Pelli, S. Soria, and G. Righini, "Spherical whispering-gallery-mode microresonators," Laser Photon. Rev. **4**(3), 457-482 (2010).
14. S. B. Gorajoobi, G. S. Murugan, and M. N. Zervas, "Design of rare-earth-doped microbottle lasers," Opt. Express **26**(20), 26339-36354 (2018).
15. H. J. Kimble, "The quantum internet," Nature **453**, 1023-1030 (2008).
16. J. Volz, M. Scheucher, C. Junge, and A. Rauschenbeutel, "Nonlinear π phase shift for single fibre-guided photons interacting with a single resonator-enhanced atom," Nat. Photon. **8**, 965-970 (2014).
17. A. Reiserer and G. Rempe, "Cavity-based quantum networks with single atoms and optical photons," Rev. Mod. Phys. **87**(4), 1379-1418 (2015).
18. A. Yariv, "Universal relations for coupling of optical power between microresonators and dielectric waveguides," Electron. Lett. **36**(4), 321-322 (2000).
19. M. Sumetsky, "Theory of SNAP devices: basic equations and comparison with the experiment," Opt. Express **20**(20), 22537-22554 (2012).
20. M. Sumetsky, "Delay of Light in an Optical Bottle Resonator with Nanoscale Radius Variation: Dispersionless, Broadband, and Low Loss," Phys. Rev. Lett. **111**, 163901 (2013).
21. M. Sumetsky, "Microscopic optical buffering in a harmonic potential," Sci. Rep. **5**, 18569 (2015).
22. S. V. Suchkov, M. Sumetsky, and A. A. Sukhorukov, "Frequency comb generation in SNAP bottle resonators," Opt. Lett. **42**(11), 2149-2152 (2017).
23. M. Sumetsky and J. Fini, "Surface nanoscale axial photonics," Opt. Express **19**(27), 26470-26485 (2011).
24. M. Sumetsky, "Nanophotonics of optical fibers," Nanophotonics **2**(5-6), 393-406 (2013).
25. L. Ding, C. Belacel, S. Ducci, G. Leo, and I. Favero, "Ultralow loss single-mode silica tapers manufactured by a microheater," Appl. Opt. **49**(13), 2441-2445 (2010).
26. M. Crespo-Ballesteros, Y. Yang, N. Toropov, and M. Sumetsky, "Four-port SNAP microresonator device," Opt. Lett. **44**(14), 3498-3501 (2019).
27. M. Sumetsky, "A SNAP coupled microresonator delay line," Opt. Express **21**(13), 15268-15279 (2013).
28. M. Sumetsky, Y. Dulashko, J. M. Fini, A. Hale, and J. W. Nicholson, "Probing optical microfiber nonuniformities at nanoscale," Opt. Lett. **31**(16), 2393-2395 (2006).
29. Y. Colombe, T. Steinmetz, G. Dubois, F. Linke, D. Hunger, and J. Reichel, "Strong atom–field coupling for Bose–Einstein condensates in an optical cavity on a chip," Nature **450**, 272-276 (2007).